\documentstyle[epsfig,aps,twocolumn,amssymb,floats,tighten]{revtex}

\def\Journal#1#2#3#4{{#1} {\bf #2}, #3 (#4)}
\def\NIMR{Nucl. Instr. and Meth. in Phys. Res. A}
\def\NPA{Nucl. Phys. A}
\def\PLB{Phys. Lett.  B}
\def\PRL{Phys. Rev. Lett.}
\def\PRC{Phys. Rev. C}

\def\PR{Phys. Rep.} 

\def\EPJA{Eur. J. Phys. A}

\def\ARNPS{Ann. Rev. Nucl. Part. Sc.}

\def\be{\begin{equation}}
\def\ee{\end{equation}}

\newcommand{\pt}{$p_t^{(0)}$}

\newcommand{\buc}     {$\rm^{1}$}
\newcommand{\bud}     {$\rm^{2}$}
\newcommand{\clt}     {$\rm^{3}$}
\newcommand{\gsi}     {$\rm^{4}$}
\newcommand{\dre}     {$\rm^{5}$}
\newcommand{\hei}     {$\rm^{6}$}
\newcommand{\ite}     {$\rm^{7}$}
\newcommand{\kur}     {$\rm^{8}$}
\newcommand{\kor}     {$\rm^{9}$}
\newcommand{\ire}     {$\rm^{10}$}
\newcommand{\war}     {$\rm^{11}$}
\newcommand{\zag}     {$\rm^{12}$}

\title{Differential directed flow in Au+Au collisions}

\vspace{0.5cm}
\author{
A.\,Andronic\gsi$^{\rm ,}$\buc,
W.\,Reisdorf\gsi,
J.P.\,Alard\clt, 
V.\,Barret\clt, Z.\,Basrak\zag, N.\,Bastid\clt, A.\,Bendarag\clt, G.\,Berek\bud,
R.\,\v{C}aplar\zag, P.\,Crochet\clt, 
A.\,Devismes\gsi, P.\,Dupieux\clt, M.\,D\v{z}elalija\zag,
C.\,Finck\gsi, Z.\,Fodor\bud,
A.\,Gobbi\gsi, Yu.\,Grishkin\ite,
O.N.\,Hartmann\gsi, N.\,Herrmann\hei, K.D.\,Hildenbrand\gsi, B.\,Hong\kor,
J.\,Kecskemeti\bud, Y.J.\,Kim\kor,
M.\,Kirejczyk\war, P.\,Koczon\gsi, M.\,Korolija\zag, R.\,Kotte\dre, 
T.\,Kress\gsi, R.\,Kutsche\gsi,
A.\,Lebedev\ite, Y.\,Leifels\gsi,
W.\,Neubert\dre,
D.\,Pelte\hei, M.\,Petrovici\buc,
F.\,Rami\ire,
B.\,de Schauenburg\ire, D.\,Sch\"ull\gsi, Z.\,Seres\bud,
B.\,Sikora\war, K.S.\,Sim\kor, V.\,Simion\buc, K.\,Siwek-Wilczy\'nska\war, 
V.\,Smolyankin\ite, M.R.\,Stockmeier\hei, G.\,Stoicea\buc, 
P.\,Wagner\ire, 
K.\,Wi\'{s}niewski\gsi, D.\,Wohlfarth\dre,
I.\,Yushmanov\kur,
A.\,Zhilin\ite \\~
(FOPI Collaboration)
} 
\vspace{0.25cm}
\address{
\buc~National Institute for Physics and Nuclear Engineering, Bucharest, Romania\\
\bud~KFKI Research Institute for Particle and Nuclear Physics, Budapest, Hungary\\
\clt~Laboratoire de Physique Corpusculaire, IN2P3/CNRS,
and Universit\'{e} Blaise Pascal, Clermont-Ferrand, France\\
\gsi~Gesellschaft f\"ur Schwerionenforschung, Darmstadt, Germany\\
\dre~Forschungszentrum Rossendorf, Dresden, Germany\\
\hei~Physikalisches Institut der Universit\"at Heidelberg, Heidelberg, Germany\\
\ite~Institute for Theoretical and Experimental Physics, Moscow, Russia\\
\kur~Kurchatov Institute, Moscow, Russia \\
\kor~Korea University, Seoul, South Korea\\
\ire~Institut de Recherches Subatomiques, IN2P3-CNRS, Universit\'e
Louis Pasteur, Strasbourg, France \\
\war~Institute of Experimental Physics, Warsaw University, Poland\\
\zag~Rudjer Boskovic Institute, Zagreb, Croatia\\ 
} 

\begin{document}

\maketitle

\begin{abstract}
We present experimental data on directed flow in semi-central Au+Au 
collisions at incident energies from 90 to 400$\cdot$A~MeV.
For the first time for this energy domain, the data are presented in a 
transverse momentum differential way.
We study the first order Fourier coefficient $v_1$ for different particle 
species and establish a gradual change of its patterns as a function of
incident energy and for different regions in rapidity.
\end{abstract}

\vspace{2mm}

\vspace{3mm}
The main motivation for the study of relativistic heavy ion collisions
is to learn about the equation of state (EoS) of nuclear matter \cite{sto86}. 
The collective phenomena  \cite{sto80} (see ref. \cite{wil97,her99} for 
recent reviews) were proposed as probes that preserve information about 
the transient hot and compressed state created in such collisions.
The (in-plane) directed flow (also called sidewards flow or transverse flow) 
was much studied, both experimentally
\cite{dos86,zha90,ram95,par95,cro97a,cro97b,ram99,mag00}
and theoretically 
\cite{aic87,gal90,bla91,aic91,jae92,pan93,ono93,fuc96,li99}.
The dependences of this phenomenon on incident energy, particle type and
centrality have been determined experimentally (see ref.~\cite{wil97,her99} 
and references therein), particularly for the Au+Au system.
The balance energy, $E_{bal}$, which is the energy of ``disappearance of flow'',
was recently directly measured for Au+Au collisions \cite{mag00}.
It was found to be around 40$\cdot$A~MeV, somewhat lower than previous 
extrapolations \cite{zha90,par95,cro97a}.
The main features of the directed flow have been reproduced by the 
theoretical models, but a final conclusion on the EoS has not yet been 
achieved. As pointed out early on \cite{aic87,gal90,bla91}, the momentum 
dependent interactions (MDI) play a crucial role in the determination 
of the EoS. Together with the (in-medium) nucleon-nucleon cross section 
($\sigma_{nn}$), MDI has a marked influence on the directed flow 
\cite{pan93,ono93,li99}. Moreover, consistency is needed in deriving EoS 
together with both MDI and $\sigma_{nn}$ \cite{jae92}.

All the above-mentioned results have been obtained in a transverse momentum 
($p_t$) integrated way, usually by studying the average in-plane transverse 
momentum, $\langle p_x\rangle$, as a function of rapidity, $y$.
The analysis of flow in terms of $p_t$ dependence of the first order Fourier 
coefficient, which we shall call ``differential directed flow'' (DDF),
has been proposed by Pan and Danielewicz \cite{pan93}, who found it sensitive
to MDI.
Li and Sustich have studied the sensitivity of DDF to both EoS and 
nucleon-nucleon cross section ($\sigma_{nn}$) around the balance energy 
\cite{li99}.
They found complex patterns of DDF, a vanishing slope of the 
$\langle p_x\rangle-y$ distribution being the result of averaging over
transverse momenta.
The DDF was extensively exploited at AGS energies both 
experimentally, by the E877 collaboration \cite{bar99}, and theoretically 
\cite{li96}. It was found to be sensitive to the characteristics of the 
transverse expansion and on the EoS, especially at high $p_t$ \cite{li96}.
For our energy domain, apart from its importance in revealing the dynamics 
of the heavy ion collisions, the differential flow could ultimately help 
towards pinning down the stiffness of the EoS of hot and dense nuclear matter 
formed in such collisions \cite{pan93,dan00}.

In this Rapid Communication we present experimental data on directed flow for 
Au+Au collisions at incident energies of 90, 120, 150, 250 and 400$\cdot$A~MeV.
Along with the integrated distributions, the data are presented in a $p_t$ 
differential way for the first time for this energy domain.
The DDF is analyzed for different particle species and as a function of 
rapidity.

The data have been measured with a wide phase-space coverage using the 
FOPI detector \cite{gob93} at GSI Darmstadt. 
The reaction products were identified by charge (Z) in the forward Plastic 
Wall (PW) at 1.2$^\circ <\Theta_{lab}<$~30$^\circ$ using time-of-flight and 
specific energy loss. In the Central Drift Chamber (CDC), covering
34$^\circ <\theta_{lab}<$~145$^\circ$, the particle identification is 
obtained using magnetic rigidity and the energy loss. 
For more details on the detector configuration for this experiment see 
ref.~\cite{and00}.

For the centrality selection we used the charged particles
multiplicities, classified into five bins. 
The variable $E{rat}=\sum_{i}E_{\perp,i}/\sum_{i}E_{\parallel,i}$ 
(the sums run over the transverse and longitudinal c.m. kinetic energy 
components of all the products detected in an event) has been additionally 
used for a better selection of the most central collisions.
The geometric impact parameters interval for our centrality 
region M4, for which the results are presented here, is 2.0--5.3 fm.
We note that in this bin the directed flow exhibits a maximum as a
function of centrality \cite{wil97,ram99}.

To compare different incident energies, we use normalized 
center-of-mass (c.m.) transverse momentum (per nucleon) and rapidity, 
defined as:
\be p_t^{(0)}=(p_t/A)/(p_P^{cm}/A_P), \quad y^{(0)}=(y/y_P)^{cm}\label{eq-1}\ee
where the subscript $P$ denotes the projectile.

The reaction plane has been reconstructed event-by-event using the transverse
momentum method \cite{dan85}.
All particles in an event have been used for the reaction plane 
reconstruction, excluding the particle-of-interest to prevent 
autocorrelations and a window around midrapidity ($|y^{(0)}|<$ 0.3) 
to improve the resolution.
The correction of the extracted values due to the reconstructed reaction 
plane fluctuations has been done using the recipe of Ollitrault \cite{oli97}. 
The correction factors, 1/$\langle\cos\Delta\phi\rangle$, where $\Delta\phi$ 
is the resolution of the reaction plane azimuth, are presented in 
Table~\ref{tab-1} for the five incident energies, along with the projectile 
momenta per nucleon in the c.m. system, $p_P^{cm}/A_P$.

\begin{table}[hbt]
\caption{Projectile momentum (per nucleon) in the center of mass and the 
correction factor for the reaction plane resolution.}\label{tab-1}
\begin{tabular}{lccccc}
E (A~MeV) ~~~~~~~~~    & 90   & 120  & 150  & 250  & 400  \\ \hline
$p_P^{cm}/A_P$ (MeV/c) & 205& 237& 265& 342& 433\\ 
1/$\langle\cos\Delta\phi\rangle$ &1.51&1.18 &1.10 &1.05 &1.04 \\
\end{tabular}
\end{table}

An important ingredient in the present analysis is the correction for 
distortions due to multiple hit losses. 
As an example, in case of PW, despite its good granularity (512 independent 
modules \cite{gob93}), average multiple hit probabilities of up to about 
9\% at 400$\cdot$A~MeV are registered for the multiplicity bin M4. 
Because of the directed flow, the average number of particles detected over 
the full PW subdetector is 2 times higher than out of the reaction plane.
This leads to an underestimation of the directed flow and needs to be 
taken into account. We developed a correction procedure based on the 
experimental data, by exploiting the DDF left-right symmetry with respect 
to midrapidity. 
The correction depends on \pt and incident energy. 
It reaches 12\% for the energy of 400$\cdot$A~MeV and is almost negligible 
at 90$\cdot$A~MeV.
The procedure was checked and validated using IQMD (Isospin Quantum Molecular 
Dynamics \cite{aic91}) events passed through a full GEANT simulation of the 
detector.

\begin{figure}[hbt] 
\centering\mbox{\epsfig{file=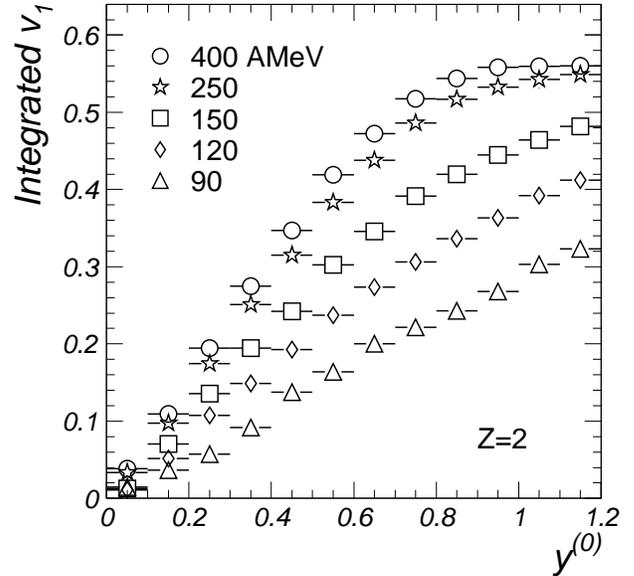,width=.45\textwidth}}
\caption{Integrated $v_1$ values as a function of rapidity for five 
incident energies, M4 centrality bin, for Z=2 particles.}
\label{fig-1} 
\end{figure} 

We characterize the directed flow by the first order Fourier 
coefficient: 
$ v_1=\langle\cos(\phi)\rangle$
where $\phi$ is the angle with respect to the reaction plane. 
In Fig.~\ref{fig-1} we present for Z=2 particles the
values of the $v_1$ coefficient integrated over transverse momentum.
These distributions are for the forward hemisphere and were obtained 
by combining the information of both PW and CDC.
Note that, for the phase space coverage of CDC, Z=2 sample comprises only 
alpha particles. 
We mention that the slopes at midrapidity of the equivalent 
$\langle p_x\rangle-y^{(0)}$ distributions are in perfect agreement 
with the FOPI results from Phase I data \cite{cro97a}.

Two important features can be emphasized from the distributions presented 
in Fig.~\ref{fig-1}.
First, we are in a region where the flow is strongly increasing as a function 
of incident energy \cite{cro97a}, as a result of stronger compression. 
This is contrary to expectations based on ideal hydrodynamics 
\cite{bon87} and has been related to a possible liquid-gas phase 
transition which breaks the scaling behaviour \cite{bon87,cro97a}.
Second, the shape as a function of rapidity is evolving with the incident 
energy. For higher energies, a saturation of the $v_1$ coefficient is seen 
for the projectile rapidities. 
Weighted with the (rapidity dependent) transverse momentum spectra
it leads to the well known so-called ``S curve'' in the 
$\langle p_x\rangle-y$ distributions \cite{dos86,ram95,par95,cro97a,ram99}
which was interpreted as a result of ``bounce-off'' from the spectators.
At lower energies, the steady evolution of $v_1$ as a function of $y$ 
points to a poorer separation of the participant and spectator contributions,
probably as a result of a relatively stronger Fermi motion.
Intuitively, longer passing times might as well allow for larger cross-talk 
and ``equilibration'' between participant and spectators.

\begin{figure}[hbt] 
\centering\mbox{\epsfig{file=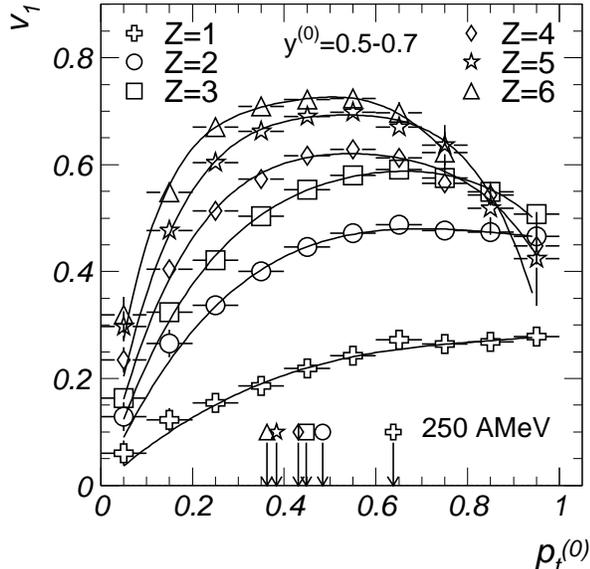, width=.43\textwidth}}
\caption{The dependence of the $v_1$ coefficient on transverse momentum for
250$\cdot$A~MeV incident energy, M4 centrality bin, for different particles.
The lines are polynomial fits to guide the eye.
The arrows mark the values of the average \pt for each particle.}
\label{fig-2x} 
\end{figure} 

In Fig.~\ref{fig-2x} we show the differential flow for particles with charge 
Z=1 to Z=6 for the incident energy E=250$\cdot$A~MeV.
We selected the rapidity window 0.5$<y^{(0)}<$0.7, corresponding to the
maximum of the directed flow in the $\langle p_x\rangle$-$y$ representation.
From here on, to keep a clean particle selection on Z, we limit 
our $p_t$ domain to the PW acceptance (which translates in a 
rapidity-dependent upper \pt limit).
The error bars on $v_1$ include both statistical and systematical errors, 
the latter coming from the apparatus acceptance. 
The horizontal bars indicate the \pt bin width.
The arrows, labeled by the corresponding symbols, are indicating the average 
values of \pt for the fragments analyzed.
The lines here and in the following figures are polynomial fits to guide 
the eye.
For the chosen rapidity window the multiplicities of the fragments up to
Z=6 relative to Z=1 are in the ratios 1000/380/68/20/14/8.
We first remark that the flow of heavier fragments is larger, a fact 
established in an integrated way by many experiments \cite{par95,ram99}
and interpreted as a signature of a collective phenomenon \cite{oli97}.
As was demonstrated in ref.~\cite{ram99}, this is the result of 
the interplay of the thermal and collective motion. For the lighter particles 
the thermal contribution is higher relative to the collective one, leading 
to a lower ``apparent'' flow.
One can see in Fig.~\ref{fig-2x} that not only the magnitude is increasing,
but also that the pattern of $v_1$ changes for heavier particles, for 
which a maximum develops at steadily lower scaled momenta.
The increase of the anisotropy as a function of $p_t$ is very 
rapid and compatible with a linear one only in a very narrow $p_t$ region. 
Note that at AGS energies a linear dependence of $v_1$ on $p_t$ was 
established experimentally for different particle species over a broad 
$p_t$ interval \cite{bar99}.

\begin{figure}[hbt]
\centering\mbox{\epsfig{file=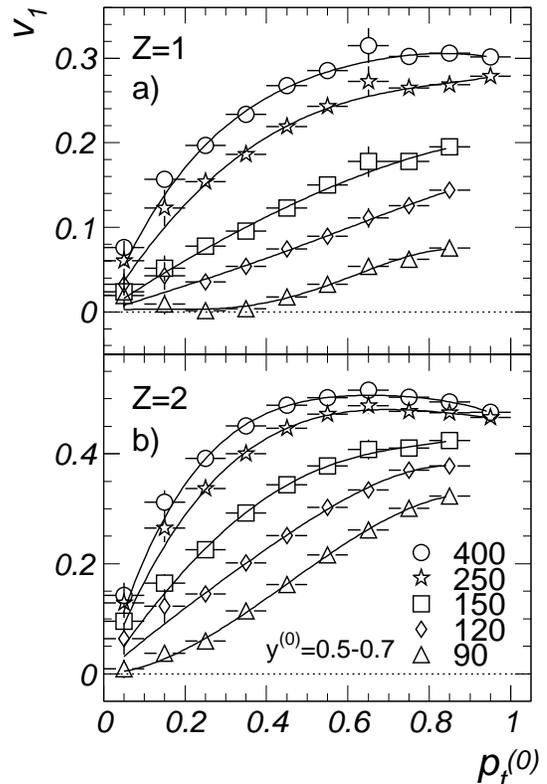, width=.39\textwidth}}
\caption{The dependence of the $v_1$ coefficient on transverse momentum for
five incident energies, M4 centrality bin, for Z=1 and Z=2 particles.}
\label{fig-2} 
\end{figure} 

Fig.~\ref{fig-2} presents, for our five incident energies, the 
dependence of the $v_1$ coefficient on the normalized transverse momentum 
for Z=1 and Z=2 particles in the rapidity window 0.5$<y^{(0)}<$0.7. 
The two particle species show different patterns of DDF as
function of incident energy.
For the lowest incident energy, 90$\cdot$A~MeV, while for Z=1 particles 
the flow is close to zero for the low \pt part and shows a later onset, 
in the case of Z=2 fragments one can notice a steady increase of $v_1$ 
as a function of \pt. 
Although with smaller magnitude (as we are well above $E_{bal}$), the
Z=1 case is similar to the behavior predicted by Li and Sustich 
at $E_{bal}$~\cite{li99}.
The coexistence of two flow patterns, attractive and repulsive, was explained 
in ref.~\cite{li99} by the fact that the low $p_t$ particles suffer more 
the influence of the attractive mean field, leading to a negative $v_1$,
while at high $p_t$ the repulsive nucleon-nucleon scatterings are responsible
for the positive flow.
The particular feature of DDF seen at 90$\cdot$A~MeV is rapidly changing 
as a function of incident energy, the increase of $v_1$ as a function of $p_t$ 
becoming gradually steeper for higher energies.

For both particle species there is a gradual development of a limiting 
value of DDF at high \pt as the incident energy increases. Part of these 
high-$p_t$ particles could have been emitted at a pre-equilibrium stage, 
therefore not reaching the maximum compression stage of the reaction. 
For the corresponding rapidity window, the average transverse momentum 
of Z=1 particles has a more pronounced variation compared to Z=2 (see 
Table~\ref{tab-2}), presumably as a result of a different participation
of different particle types in the reaction dynamics at these energies 
\cite{pet95}.
Also, Z=2 particles can originate to a large extent from secondary decays 
of heavier fragments \cite{ono93}.
It would be important to study these details using transport models.

\begin{table}[hbt]
\caption{Average normalized transverse momentum for particles with Z=1 and
Z=2 for the rapidity window $y^{(0)}$=0.5-0.7.}\label{tab-2}
\begin{tabular}{lccccc}
Energy (A~MeV) ~~~~    & 90   & 120  & 150  & 250  & 400  \\ \hline
$\langle p_t^{(0)}\rangle$ ~~ Z=1 & 0.74& 0.71& 0.67& 0.63& 0.60\\
$\langle p_t^{(0)}\rangle$ ~~ Z=2 & 0.51& 0.50& 0.49& 0.48& 0.48 \\ 
\end{tabular}
\end{table}

\begin{figure}[hbt]
\centering\mbox{\epsfig{file=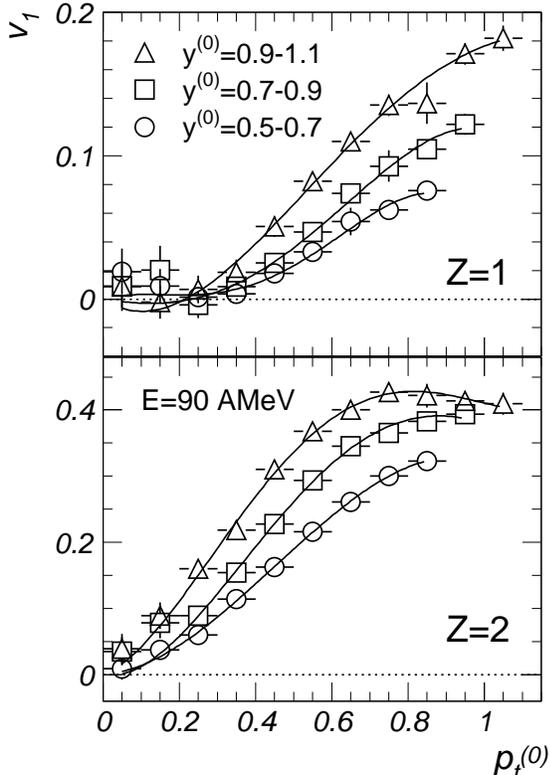, width=.4\textwidth}}
\caption{$v_1$ as a function of transverse momentum in three windows in 
rapidity for particles with Z=1 and Z=2 for the incident energy of 
90$\cdot$A~MeV.}
\label{fig-3}
\end{figure} 

From our differential flow values at 90$\cdot$A~MeV one can infer that 
$E_{bal}$ is not a single-defined value but is dependent on particle type 
and on transverse momentum. 
We have observed similar trends in what concerns $E_{tran}$, the energy of 
the transition from in-plane to out-of-plane azimuthal enhancement at 
midrapidity \cite{and00}.
As $E_{bal}$ is decreasing as a function of $p_t$,
its integrated value cannot be compared among various experiments unless
their different thresholds are taken into account. 
Moreover, $E_{bal}$ can depend on rapidity as well, as shown in 
Fig.~\ref{fig-3}, where the DDF is presented for Z=1 and Z=2 
particles for three windows in rapidity for the incident energy of 
90$\cdot$A~MeV.  
Notice again the difference between the two particle species in what
concerns the patterns of the flow and that at projectile rapidities
(triangles) the anisotropy of Z=2 particles already develops a maximum,
in contrast to Z=1 particles.

\begin{figure}[hbt]
\centering\mbox{\epsfig{file=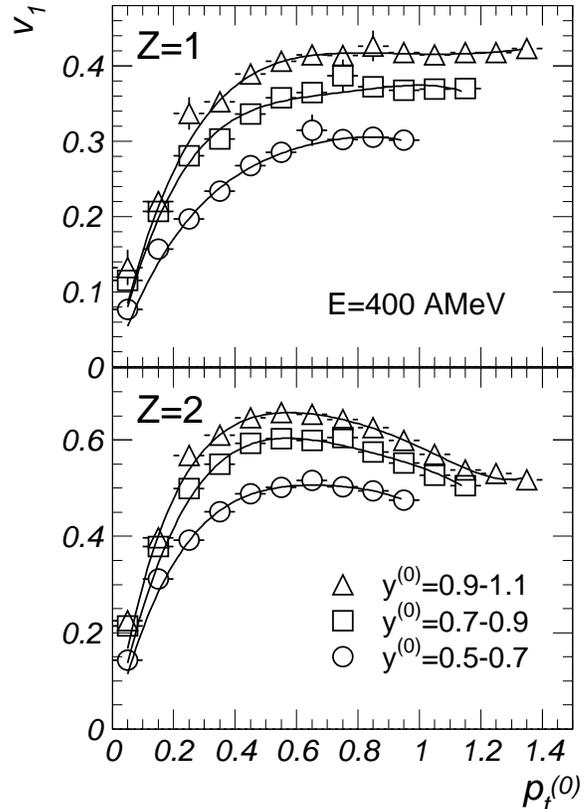, width=.42\textwidth}}
\caption{Same as Fig.~\ref{fig-3}, but for the incident energy of 
400$\cdot$A~MeV.}
\label{fig-4}
\end{figure} 

While at our lowest incident energy of 90$\cdot$A~MeV the DDF 
may help towards pinning down the in-medium nucleon-nucleon cross section 
\cite{li99}, at higher energies it may bear significance on the MDI 
\cite{pan93} and ultimately on the EoS. 
The rapidity dependence of the DDF at the incident energy of 
400$\cdot$A~MeV is presented in Fig.~\ref{fig-4} for Z=1 and Z=2 particles.  
Again, at all rapidities there is a clear difference in shape between the 
two particle species. 
For Z=2 particles notice that at projectile rapidities ($y^{(0)}$=1)
the anisotropy is highest (see also Fig.~\ref{fig-1}), while the 
$\langle p_x \rangle$ values would already show a decreasing trend
\cite{dos86,ram95,par95,cro97a,ram99}.

For both incident energies, Fig.~\ref{fig-3} and ~\ref{fig-4}, one can see 
that the shape of the distributions varies as a function of rapidity,
giving additional support for the influence of the dynamics of
the collision on the flow values. 
Also, the maximum of $v_1$ for Z=2 (for $y^{(0)}$=0.9-1.1) for the energy of 
400$\cdot$A~MeV occurs at a different value of transverse momentum
compared to 90$\cdot$A~MeV, either in terms of scaled or not scaled 
momenta.

In summary, we have shown first experimental results on differential
directed flow in heavy ion collisions at incident energies from 90 to 
400$\cdot$A~MeV. 
Our $p_t$ differential study allows us to show that not only is the flow
larger, but also its pattern is different for heavier particles. 
In addition, it has a complex evolution as a function of beam energy.
The differential flow pattern of Z=1 and Z=2 particles was studied 
as well for different regions in rapidity.

It is expected that the DDF will help to clear the difficulties (and sometimes 
inconsistencies) noticed in previous studies devoted to directed flow using 
comparisons with transport models, either of QMD-type 
\cite{aic87,aic91,jae92,par95,cro97b} 
or from the Boltzmann-Uehling-Uhlenbeck (BUU) family \cite{gal90,pan93,fuc96}.
Present comparisons of the experimental differential directed flow to 
BUU \cite{gai01} or IQMD \cite{and01} models show sensitivities to the EoS.
But, if the EoS is to be extracted from such types of comparisons, more effort
is needed, especially in disentangling the contributions of momentum dependent
interactions and in-medium nucleon-nucleon cross sections from 
the EoS itself. 
It is in the energy range of the present study that both MDI and $\sigma_{nn}$
have a very important contribution and cannot at all be neglected.
Along with the centrality dependence of directed flow 
\cite{aic87,pan93,cro97b} and with the differential elliptic flow \cite{dan00},
the DDF can contribute to a good extent in fixing these quantities 
\cite{pan93,li99}.
Properly incorporating the fragment production \cite{ono93} and 
a consistent treatment of the momentum dependent interactions 
\cite{jae92,ono93,fuc96} and in-medium nucleon-nucleon cross sections
are essential prerequisites for this task.

This work has been supported in part by the German BMBF under contracts 
RUM-005-95, POL-119-95, UNG-021-96 and RUS-676-98 and by the Deutsche 
Forschungsgemeinschaft (DFG) under projects 436 RUM-113/10/0, 
436 RUS-113/143/2 and 446 KOR-113/76/0,
Support has also been received from the Polish State Committee of Scientific 
Research, KBN, from the Hungarian OTKA under grant T029379, from the 
Korea Research Foundation under contract No. 1997-001-D00117,
from the agreement between GSI and CEA/IN2P3 and from the PROCOPE Program
of DAAD.

\end{document}